\documentclass[twocolumn,showpacs,prl,superscriptaddress]{revtex4}          
\def\sNN{\mbox{$\sqrt{s_{_{NN}}}$}}   
\newcommand{ \be }{\begin{equation}}       
\newcommand{ \ee }{\end{equation}}       
\newcommand{ \bea }{\begin{eqnarray}}       
\newcommand{ \eea }{\end{eqnarray}}

\newcommand{ \mean }[1]{\left\langle #1 \right\rangle}   
\newcommand{ \etal }{{\it et al.}}   
\usepackage{color}

\usepackage{graphicx} 
\usepackage{fixmath} 
\usepackage{dcolumn} 
\usepackage{bm} 
\usepackage{multirow}
\usepackage{ulem} 
\usepackage{lineno}
\modulolinenumbers[2]

\begin{document}          
\title{       
Beam-energy dependence of charge separation along the magnetic field in Au+Au collisions at RHIC
} 
 
 
\affiliation{AGH University of Science and Technology, Cracow, Poland}
\affiliation{Argonne National Laboratory, Argonne, Illinois 60439, USA}
\affiliation{University of Birmingham, Birmingham, United Kingdom}
\affiliation{Brookhaven National Laboratory, Upton, New York 11973, USA}
\affiliation{University of California, Berkeley, California 94720, USA}
\affiliation{University of California, Davis, California 95616, USA}
\affiliation{University of California, Los Angeles, California 90095, USA}
\affiliation{Universidade Estadual de Campinas, Sao Paulo, Brazil}
\affiliation{Central China Normal University (HZNU), Wuhan 430079, China}
\affiliation{University of Illinois at Chicago, Chicago, Illinois 60607, USA}
\affiliation{Cracow University of Technology, Cracow, Poland}
\affiliation{Creighton University, Omaha, Nebraska 68178, USA}
\affiliation{Czech Technical University in Prague, FNSPE, Prague, 115 19, Czech Republic}
\affiliation{Nuclear Physics Institute AS CR, 250 68 \v{R}e\v{z}/Prague, Czech Republic}
\affiliation{Frankfurt Institute for Advanced Studies FIAS, Germany}
\affiliation{Institute of Physics, Bhubaneswar 751005, India}
\affiliation{Indian Institute of Technology, Mumbai, India}
\affiliation{Indiana University, Bloomington, Indiana 47408, USA}
\affiliation{Alikhanov Institute for Theoretical and Experimental Physics, Moscow, Russia}
\affiliation{University of Jammu, Jammu 180001, India}
\affiliation{Joint Institute for Nuclear Research, Dubna, 141 980, Russia}
\affiliation{Kent State University, Kent, Ohio 44242, USA}
\affiliation{University of Kentucky, Lexington, Kentucky, 40506-0055, USA}
\affiliation{Korea Institute of Science and Technology Information, Daejeon, Korea}
\affiliation{Institute of Modern Physics, Lanzhou, China}
\affiliation{Lawrence Berkeley National Laboratory, Berkeley, California 94720, USA}
\affiliation{Massachusetts Institute of Technology, Cambridge, Massachusetts 02139-4307, USA}
\affiliation{Max-Planck-Institut f\"ur Physik, Munich, Germany}
\affiliation{Michigan State University, East Lansing, Michigan 48824, USA}
\affiliation{Moscow Engineering Physics Institute, Moscow Russia}
\affiliation{National Institute of Science Education and Research, Bhubaneswar 751005, India}
\affiliation{Ohio State University, Columbus, Ohio 43210, USA}
\affiliation{Old Dominion University, Norfolk, Virginia 23529, USA}
\affiliation{Institute of Nuclear Physics PAN, Cracow, Poland}
\affiliation{Panjab University, Chandigarh 160014, India}
\affiliation{Pennsylvania State University, University Park, Pennsylvania 16802, USA}
\affiliation{Institute of High Energy Physics, Protvino, Russia}
\affiliation{Purdue University, West Lafayette, Indiana 47907, USA}
\affiliation{Pusan National University, Pusan, Republic of Korea}
\affiliation{University of Rajasthan, Jaipur 302004, India}
\affiliation{Rice University, Houston, Texas 77251, USA}
\affiliation{University of Science and Technology of China, Hefei 230026, China}
\affiliation{Shandong University, Jinan, Shandong 250100, China}
\affiliation{Shanghai Institute of Applied Physics, Shanghai 201800, China}
\affiliation{SUBATECH, Nantes, France}
\affiliation{Temple University, Philadelphia, Pennsylvania 19122, USA}
\affiliation{Texas A\&M University, College Station, Texas 77843, USA}
\affiliation{University of Texas, Austin, Texas 78712, USA}
\affiliation{University of Houston, Houston, Texas 77204, USA}
\affiliation{Tsinghua University, Beijing 100084, China}
\affiliation{United States Naval Academy, Annapolis, Maryland, 21402, USA}
\affiliation{Valparaiso University, Valparaiso, Indiana 46383, USA}
\affiliation{Variable Energy Cyclotron Centre, Kolkata 700064, India}
\affiliation{Warsaw University of Technology, Warsaw, Poland}
\affiliation{University of Washington, Seattle, Washington 98195, USA}
\affiliation{Wayne State University, Detroit, Michigan 48201, USA}
\affiliation{Yale University, New Haven, Connecticut 06520, USA}
\affiliation{University of Zagreb, Zagreb, HR-10002, Croatia}

\author{L.~Adamczyk}\affiliation{AGH University of Science and Technology, Cracow, Poland}
\author{J.~K.~Adkins}\affiliation{University of Kentucky, Lexington, Kentucky, 40506-0055, USA}
\author{G.~Agakishiev}\affiliation{Joint Institute for Nuclear Research, Dubna, 141 980, Russia}
\author{M.~M.~Aggarwal}\affiliation{Panjab University, Chandigarh 160014, India}
\author{Z.~Ahammed}\affiliation{Variable Energy Cyclotron Centre, Kolkata 700064, India}
\author{I.~Alekseev}\affiliation{Alikhanov Institute for Theoretical and Experimental Physics, Moscow, Russia}
\author{J.~Alford}\affiliation{Kent State University, Kent, Ohio 44242, USA}
\author{C.~D.~Anson}\affiliation{Ohio State University, Columbus, Ohio 43210, USA}
\author{A.~Aparin}\affiliation{Joint Institute for Nuclear Research, Dubna, 141 980, Russia}
\author{D.~Arkhipkin}\affiliation{Brookhaven National Laboratory, Upton, New York 11973, USA}
% ...

\collaboration{STAR Collaboration}\noaffiliation

\begin{abstract}     
Local parity-odd domains are theorized to form inside a Quark-Gluon-Plasma (QGP) which has been produced in high-energy heavy-ion 
collisions.
The local parity-odd domains manifest themselves as charge separation along the magnetic field axis via 
the chiral magnetic effect (CME). 
The experimental observation of charge separation has previously been reported for 
heavy-ion collisions at the top RHIC energies.
In this paper, we present the results of the beam-energy dependence of the charge correlations 
in Au+Au collisions at midrapidity for center-of-mass energies of 7.7, 11.5, 19.6, 27, 39 
and 62.4 GeV from the STAR experiment. 
After background subtraction, the signal gradually reduces with decreased beam energy, 
and tends to vanish by 7.7 GeV. 
This implies the dominance of hadronic interactions over partonic ones at lower collision energies.
\end{abstract} 
 
  
\maketitle  
The strong interaction is parity even at vanishing temperature and isospin density~\cite{WhittenVafa:1984},
but parity could be violated locally in microscopic domains in QCD at finite temperature
as a consequence of topologically non-trivial configurations of gauge fields~\cite{Lee1,Lee2}.
The Relativistic Heavy Ion Collider (RHIC) provides a good opportunity
to study such parity-odd ($\cal P$-odd) domains, where the local imbalance of
chirality results from the interplay of these topological configurations with
the hot, dense and deconfined Quark-Gluon-Plasma (QGP) created in heavy-ion collisions.

\begin{figure}[t]
  \includegraphics[width=0.40\textwidth]{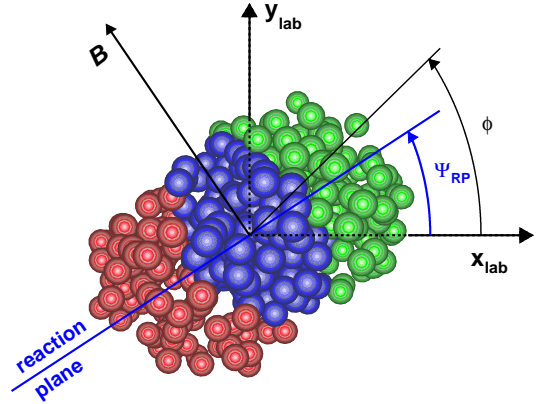}
  \caption{
    (Color online) Schematic depiction of the transverse plane for a collision of two heavy ions 
	(the left one emerging from and the right one going into the page). 
        Particles are produced in the overlap region (blue-colored nucleons). The azimuthal angles
	of the reaction plane and a produced particle used in the three-point correlator, $\gamma$, are depicted here.
}
    \label{fig:Overlap}
\end{figure}

The $\cal P$-odd domains can be manifested via the chiral magnetic effect (CME). 
In heavy-ion collisions, energetic protons (mostly spectators) produce a 
magnetic field ($B$) with a strength that peaks around
$eB \approx 10^4$ ${\rm MeV}^2$~\cite{Kharzeev}. 
The collision geometry is illustrated in Fig.~\ref{fig:Overlap}.
The strong magnetic field, coupled with the chiral asymmetry in the $\cal P$-odd domains,
induces a separation of electric charge along the direction of the magnetic field
\cite{Kharzeev,Kharzeev2,Kharzeev3,Kharzeev4,Kharzeev5,axial}.
Based on data from the STAR~\cite{STAR_LPV1,STAR_LPV2,STAR_LPV3,STAR_LPV4} and PHENIX~\cite{PHENIX_LPV1,PHENIX_LPV2} collaborations 
at RHIC and the ALICE collaboration~\cite{ALICE_LPV} at the LHC,
charge-separation fluctuations have been experimentally observed.
The interpretation of these data as an indication of the CME is still under intense discussion, 
see e.g.~\cite{dis1,dis2,STAR_LPV4} and references therein.
A study of the beam-energy dependence of the charge separation effect will shed light on
the interpretation of the data.

The magnetic field axis points in the direction that is perpendicular to the reaction plane,
which contains the impact parameter and the beam momenta.
Experimentally the charge separation is measured 
perpendicular to the reaction plane with a three-point correlator, 
$\gamma \equiv \langle \cos(\phi_1 + \phi_2 -2{\rm \Psi_{RP}}) \rangle$ ~\cite{three_point_correlation}.
In Fig.~\ref{fig:Overlap}, $\phi$ and ${\rm \Psi_{RP}}$ denote the azimuthal angles of a particle and 
the reaction plane, respectively.
In practice, we approximate the reaction plane with the ``event plane" ($\rm \Psi_{EP}$) 
reconstructed with measured particles,
and then correct the measurement for the finite event plane resolution~\cite{STAR_LPV1,STAR_LPV2,STAR_LPV3}.

This Letter reports measurements of the three-point correlator, $\gamma$, for charged particles
produced in Au+Au collisions. 8M events were analyzed at the center-of-mass energy $\sNN = 62.4$ GeV (2005), 100M at 39 GeV (2010), 
46M at 27 GeV (2011), 20M at 19.6 GeV (2011), 10M at 11.5 GeV (2010) and 4M at 7.7 GeV (2010).
Events selected with a minimum bias trigger have been sorted 
into centrality classes based on uncorrected charged particle multiplicity at midrapidity.
Charged particle tracks in this analysis were reconstructed in the STAR Time Projection Chamber (TPC)~\cite{TPC-NIM}, 
within a pseudorapidity range of $|\eta|<1$ and a transverse momentum range of $0.15 < p_T < 2$ GeV/$c$.
The centrality definition and track quality cuts are the same as in Refs.~\cite{BESv2}, unless otherwise specified.
Only events within 40 cm of the center of the detector along the beam direction were selected for data sets
at $\sNN = 19.6 - 62.4$ GeV.
This cut was 50 and 70 cm for 11.5 and 7.7 GeV collisions, respectively.
To suppress events from collisions with the beam pipe (radius 3.95 cm), only those events with a radial position 
of the reconstructed primary vertex within 2 cm were analyzed.
A cut on the distance of closest approach to the primary vertex (DCA) $< 2$ cm was also 
applied to reduce the number of weak decay tracks or secondary interactions. 
The experimental observables involved in the analysis have been corrected for the particle track reconstruction efficiency.

In an event, charge separation along the magnetic field (i.e., perpendicular to the reaction plane) may be described 
phenomenologically by a sine term in the Fourier decomposition of the charged particle azimuthal distribution,
\be
\frac{dN_{\alpha}}{d\phi} \propto 1+2v_{1}\cos(\Delta\phi)+2a_\alpha\sin(\Delta\phi)+2v_2\cos(2\Delta\phi)+...
\label{eq:FourierExp}
\ee
where $\Delta\phi = \phi - \rm{\Psi_{RP}}$, and the subscript $\alpha$ ($+$ or $-$) denotes the charge sign of particles.
Conventionally $v_1$ is called ``directed flow" and $v_2$ ``elliptic flow",
and they describe the collective motion of the produced particles~\cite{Poskanzer}.
The parameter $a$ (with $a_- = -a_+$) quantifies the $\cal P$-violating effect.
However, if spontaneous parity violation occurs,
the signs of finite $a_+$ and $a_-$ will vary from event to event,
leading to $\langle a_+ \rangle = \langle a_- \rangle = 0$.
In the expansion of the three-point correlator, $\gamma \equiv \langle \cos(\phi_1 + \phi_2 -2{\rm \Psi_{RP}}) \rangle = \langle \cos(\Delta\phi_1)\cos(\Delta\phi_2)
-\sin(\Delta\phi_1)\sin(\Delta\phi_2) \rangle$,
the second term contains the fluctuation term $-\langle a_\pm a_\pm \rangle$,
which may be non-zero when accumulated over particle pairs of separate charge combinations. 
The first term ($\langle \cos(\Delta\phi_1)\cos(\Delta\phi_2)\rangle$) in the expansion 
provides a baseline unrelated to the magnetic field.

\begin{figure}[t]
  \includegraphics[width=0.45\textwidth]{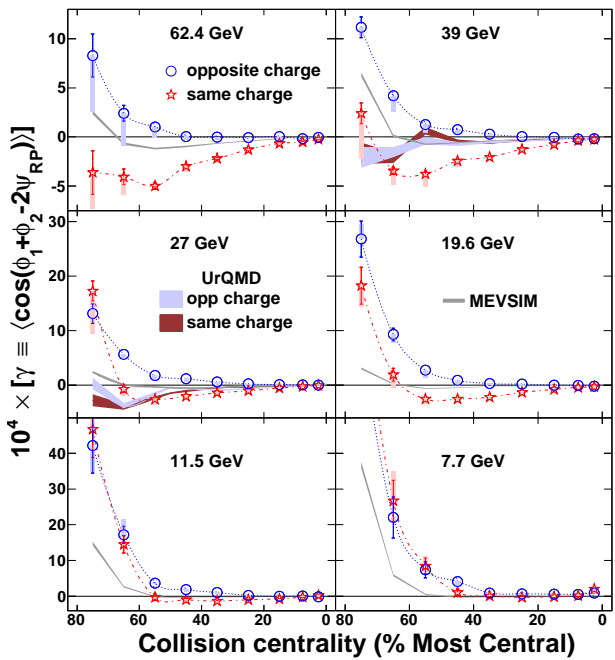}
  \caption{
    (Color online) The three-point correlator, $\gamma$, as a function of centrality for Au+Au collisions at $7.7-62.4$ GeV.
    Note that the vertical scales are different for different rows.
    The filled boxes (starting from the central values) represent one type of systematic uncertainty (as discussed in the text).
   Charge independent results from the model calculations of MEVSIM~\cite{MEVSIM} are shown as grey curves.
   $\gamma_{\rm OS}$ and $\gamma_{\rm SS}$ from UrQMD calculations~\cite{UrQMD} are also shown as shaded bands for 27 and 39 GeV.
}
    \label{fig:6panel}
\end{figure}

The reaction plane of a heavy-ion collision is not known a priori, and in practice
it is approximated with an event plane which is reconstructed from particle azimuthal distributions~\cite{Poskanzer}.
In this analysis, we exploited the large elliptic flow of charged hadrons produced at mid-rapidity
to construct the event plane angle:
\be
{\rm{\Psi_{EP}}} = \frac{1}{2}\tan^{-1} \biggl[\frac{\sum \omega_i \sin(2 \phi_i)}{\sum \omega_i \cos(2 \phi_i)} \biggr],
\ee
where $\omega_i$ is a weight for each particle $i$ in the sum~\cite{Poskanzer}.
The weight was chosen to be the $p_T$ of the particle itself,
and only particles with $p_T<2$ GeV/$c$ were used.
Although the STAR TPC has good azimuthal symmetry, small acceptance effects in the calculation of
the event plane azimuth were removed by the method of shifting~\cite{shifting}.
The observed correlations were corrected for the event plane resolution, estimated with the
correlation between two random sub-events (details in Ref.~\cite{Poskanzer}).

The event plane thus obtained from the produced particles is sometimes called ``the participant plane" since
it is subject to the event-by-event fluctuations of the initial participant nucleons~\cite{PPlane}.
A better approximation to the reaction plane could be obtained from the spectator neutron distributions detected
in the STAR zero degree calorimeters (ZDC-SMDs)~\cite{ZDC}. 
This type of event plane utilizes the directed flow of spectator neutrons
measured at very forward rapidity. 
We have measured the three point correlations using both types of reaction plane estimates and the results are
consistent with each other~\cite{STAR_LPV3}.
Other systematic uncertainties were studied extensively and discussed 
in our previous publications on the subject~\cite{STAR_LPV1,STAR_LPV2}.
All were shown to be negligible compared with the uncertainty in determining the reaction plane.
In this work, we have only used the participant plane because the efficiency of ZDC-SMDs becomes low
for low beam energies.

Figure~\ref{fig:6panel} presents the opposite-charge ($\gamma_{\rm OS}$) and same-charge ($\gamma_{\rm SS}$) correlators 
for Au+Au collisions at $\sqrt{s_{NN}} = 7.7 - 62.4$ GeV as a function of centrality (0 means the most central collisions).
In most cases, the ordering of $\gamma_{\rm OS}$ and $\gamma_{\rm SS}$ is the same
as in Au+Au (Pb+Pb) collisions at higher energies~\cite{STAR_LPV1,STAR_LPV2,STAR_LPV3,ALICE_LPV},
suggesting charge-separation fluctuations perpendicular to the reaction plane. 
As a systematic check, the charge combinations of $++$ and $--$ were always found to be consistent with each other
(not shown here).
With decreased beam energy, both $\gamma_{\rm OS}$ and $\gamma_{\rm SS}$ tend to rise up in peripheral collisions.
This feature seems to be charge independent, and can be explained by momentum conservation and elliptic flow~\cite{STAR_LPV3}.
Momentum conservation forces all produced particles, regardless of charge, to separate from each other,
while elliptic flow, a collective motion, works in the opposite sense. For peripheral collisions, the multiplicity ($N$) is small,
and momentum conservation dominates. At lower beam energies, 
$N$ also becomes smaller and hence higher values for $\gamma_{\rm OS}$ and $\gamma_{\rm SS}$.
For more central collisions where the multiplicity is large, this type of ${\cal P}$-even background 
can be estimated as $-v_2/N$~\cite{STAR_LPV3,v2N}.
In Fig.~\ref{fig:6panel}, we also show the model calculations of MEVSIM,
a Monte Carlo event generator  with an implementation of $v_2$ and momentum conservation, 
developed for STAR simulations~\cite{MEVSIM}.
The model results qualitatively describe the beam-energy dependency of the charge-independent background.

In view of the charge-independent background, the charge separation effect can be studied
via the difference between $\gamma_{\rm OS}$ and $\gamma_{\rm SS}$.
The difference $(\gamma_{\rm OS} - \gamma_{\rm SS})$ remains positive for all centralities down to the beam energy $\sim19.6$ GeV,
and the magnitude decreases from peripheral to central collisions.
Presumably this is partially owing to the reduced magnetic field
and partially owing to the more pronounced dilution effect in more central collisions. 
A dilution of the correlation is expected when there are multiple sources involved in the collision~\cite{STAR_LPV2,dilution}.
The difference between $\gamma_{\rm OS}$ and $\gamma_{\rm SS}$ approaches zero in peripheral collisions at lower energies, 
especially at 7.7 GeV, which could be understood in terms of the CME hypothesis if the formation of the QGP
becomes less likely in peripheral collisions at low beam energies~\cite{Vitaly}.

\begin{figure}[t]
  \includegraphics[width=.45\textwidth]{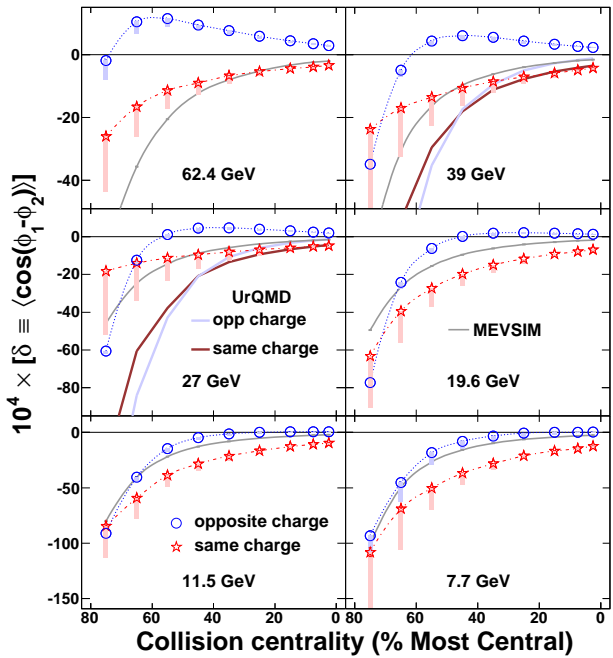}
  \caption{(Color online) The two-particle correlation as a function of centrality for Au+Au collisions at $7.7-62.4$ GeV.
    Note that the vertical scales are different for different rows.
    The filled boxes bear the same meaning as those in Fig~\ref{fig:6panel} and are described in the text.
    MEVSIM and UrQMD calculations are also shown for comparison.
        }
  \label{fig:delta}
\end{figure}

\begin{figure}[t]
  \includegraphics[width=0.41\textwidth]{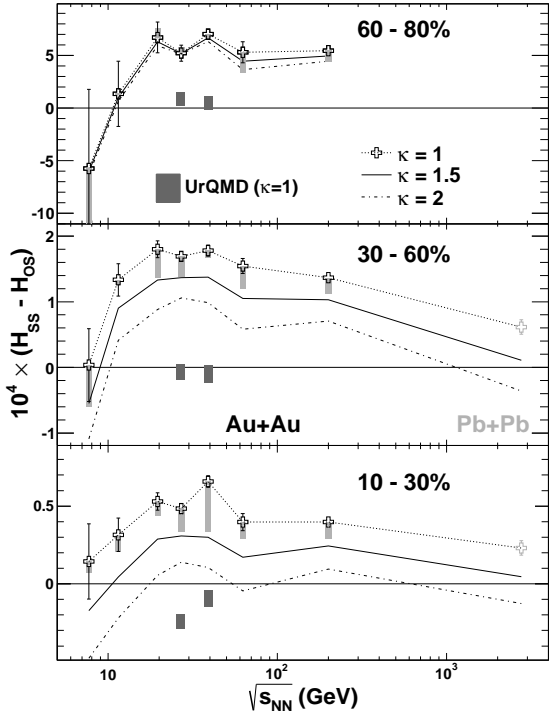}
  \caption{
    $H_{\rm SS}-H_{\rm OS}$, as a function of beam energy
         for three centrality bins in Au+Au collisions.
    The default values (dotted curves) are for $H^{\kappa=1}$, and the solid (dash-dot) curves
    are obtained with $\kappa=1.5$ ($\kappa=2$).
    For comparison, the results for Au+Au collisions at 200 GeV~\cite{STAR_LPV2} and Pb+Pb collisions at 2.76 TeV~\cite{ALICE_LPV} are also shown.
    The systematic errors of the STAR data (filled boxes) bear the same meaning as those in Fig.~\ref{fig:6panel}.
    UrQMD calculations with $\kappa=1$ are also shown as solid shaded bars for 27 and 39 GeV. 
}
    \label{fig:bes}
\end{figure}

The systematic uncertainties of $(\gamma_{\rm OS} - \gamma_{\rm SS})$ due to the analysis cuts, the track reconstruction efficiency
and the event plane determination were estimated to be approximately $10\%$, $5\%$ and $10\%$, respectively.
Overall, total systematic uncertainties are typically $15\%$,
except for the cases where $(\gamma_{\rm OS} - \gamma_{\rm SS})$ is close to zero.
Another type of uncertainty is due to quantum interference (``HBT" effects)
and final-state-interactions (Coulomb dominated)~\cite{STAR_LPV3}, 
which are most prominent for low relative momenta. 
To suppress the contributions from these effects, we applied the conditions of $\Delta p_T > 0.15$ GeV/$c$ 
and $\Delta\eta>0.15$ to the correlations, shown
as filled boxes in Figs.~\ref{fig:6panel},~\ref{fig:delta} and~\ref{fig:bes}.
The boxes start from the central values with default conditions
and end with the results with the above extra conditions on $\Delta p_T$ and $\Delta\eta$.

Interpretation of the three particle correlation result, $\gamma$, requires additional information 
such as a measurement of the two particle correlation
$\delta \equiv \mean{\cos(\phi_{1}-\phi_{2})}= \langle \cos(\Delta\phi_1)\cos(\Delta\phi_2)
+\sin(\Delta\phi_1)\sin(\Delta\phi_2) \rangle$.
The expansion of $\delta$ also contains the fluctuation term $\langle a_\pm a_\pm \rangle$
(with a sign opposite to that in $\gamma$).
Figure~\ref{fig:delta} shows $\delta$
as a function of centrality for Au+Au collisions at $7.7-62.4$ GeV.
In contrary to the CME expectation, $\delta_{\rm OS}$ is above $\delta_{\rm SS}$ in most cases, 
indicating an overwhelming background, larger than any possible CME effect. 
The background sources, if coupled with collective flow, will also contribute to $\gamma$.
Taking this into account, we express $\gamma$ and $\delta$ in the following forms,
where the unknown parameter $\kappa$, as argued in Ref.~\cite{Flow_CME}, is of the order of unity.
\bea
\gamma &\equiv& \langle \cos(\phi_1 + \phi_2 -2{\rm \Psi_{RP}}) \rangle = \kappa v_2 F - H \\
\label{eq:3}
\delta &\equiv& \langle \cos(\phi_1 -\phi_2) \rangle = F + H,
\label{eq:4}
\eea
where $H$ and $F$ are the CME and background contributions, respectively.
In Ref.~\cite{Flow_CME} $\kappa = 1$, but it could deviate from unity 
owing to a finite detector acceptance and theoretical uncertainties.
We can solve for $H$ from Eqns.~3 and \ref{eq:4},
\be
H^\kappa = (\kappa v_2 \delta - \gamma)/(1+\kappa v_2).
\ee

Figure~\ref{fig:bes} shows $H_{\rm SS}-H_{\rm OS}$ as a function of beam energy for three centrality bins in Au+Au collisions.
$v_2$ for the beam energies under study has been measured in our previous publications~\cite{BESv2}.
The default values (dotted curves) are for $H^{\kappa=1}$, and the solid (dash-dot) curves
are obtained with $\kappa=1.5$ ($\kappa=2$).
For comparison, the results for $10-60\%$ Pb+Pb collisions at 2.76 TeV are also shown~\cite{ALICE_LPV}.
The $(H_{\rm SS}-H_{\rm OS})$ curve for $\kappa =1$ suggests a non-zero charge separation effect with a weak energy dependence 
above 19.6 GeV, but the trend rapidly decreases to zero in the interval between 19.6 and 7.7 GeV.
This may be explained by the probable domination of hadronic interactions over partonic ones at low beam energies.
With increased $\kappa$, $(H_{\rm SS} - H_{\rm OS})$ decreases for all beam energies
and may even totally disappear in some case (e.g. with $\kappa \sim 2$ in $10-30\%$ collisions). 
A better theoretical estimate of $\kappa$ in the future
would enable us to extract a firmer conclusion from the data presented.

MEVSIM calculations qualitatively reproduce the charge-independent background for
both $\gamma$ and $\delta$ correlators as shown in Figs.~\ref{fig:6panel} and \ref{fig:delta}, 
but they always yield identical same-charge and opposite-charge correlations.
To further study the charge separation effect, a transport model, UrQMD~\cite{UrQMD}, was employed. 
UrQMD calculations have finite difference between same-charge and opposite-charge $\gamma$ ($\delta$) correlations,
while $H_{\rm SS}-H_{\rm OS}$ is either slightly negative or consistent with zero.
This is demonstrated for 27 and 39 GeV in Figs.~\ref{fig:6panel}, \ref{fig:delta} and \ref{fig:bes}.

In summary, an analysis of the three-point correlation between two charged particles 
and the reaction plane has been carried out for Au+Au collisions at $\sqrt{s_{NN}} = 7.7 - 62.4$ GeV. 
The general trend of the correlations ($\gamma_{\rm OS}$ and $\gamma_{\rm SS}$), as a function of centrality and beam energy,
can be qualitatively described by the model calculations of MEVSIM. 
This result indicates a large contribution from the ${\cal P-}$even background 
due to momentum conservation and collective flow.
The charge separation along the magnetic field, studied via $(H_{\rm SS} - H_{\rm OS})$, 
shows a signal with a weak energy dependence down to 19.6 GeV and then falls steeply at lower energies.
This trend may be consistent with the hypothesis of local parity violation because there should be a smaller probability 
for the CME at lower energies where the hadronic phase plays a more dominant role than the partonic phase.
A more definitive result may be obtained in the future if we can increase the statistics by a factor of ten for the low energies
and if we can reduce the uncertainty associated with determination of the value of $\kappa$. 

We thank the RHIC Operations Group and RCF at BNL, the NERSC Center at LBNL, the KISTI Center in Korea, and the Open Science Grid consortium for providing resources and support. This work was supported in part by the Offices of NP and HEP within the U.S. DOE Office of Science, the U.S. NSF, CNRS/IN2P3, FAPESP CNPq of Brazil,  the Ministry of Education and Science of the Russian Federation, NNSFC, CAS, MoST and MoE of China, the Korean Research Foundation, GA and MSMT of the Czech Republic, FIAS of Germany, DAE, DST, and CSIR of India, the National Science Centre of Poland, National Research Foundation (NRF-2012004024), the Ministry of Science, Education and Sports of the Republic of Croatia, and RosAtom of Russia.

  
\end{document}